\documentclass[conference]{IEEEtran}
\IEEEoverridecommandlockouts

\usepackage{cite}
\usepackage{amsmath,amssymb,amsfonts}
\usepackage{algorithmic}
\usepackage{graphicx}
\usepackage{textcomp}
\usepackage{xcolor}
\def\BibTeX{{\rm B\kern-.05em{\sc i\kern-.025em b}\kern-.08em
    T\kern-.1667em\lower.7ex\hbox{E}\kern-.125emX}}

\usepackage{xspace}

\usepackage[table]{xcolor} 
\usepackage{tabularx}      
\usepackage{makecell}      

\usepackage{enumitem}

\usepackage{float}
\usepackage{dirtytalk}
\usepackage{csquotes}
\usepackage{balance}

\usepackage{graphicx} 

\usepackage[normalem]{ulem}

\newcommand{\myparagraph}[1]{\textbf{#1.}}

\usepackage{flushend}
\usepackage[svgnames]{xcolor}

\usepackage{textcomp}
\usepackage[T1]{fontenc}
\usepackage{lineno}
\usepackage{booktabs}
\usepackage[table]{xcolor}
\usepackage{textcomp}

\usepackage[table]{xcolor}

\newboolean{showcomments}
\setboolean{showcomments}{true}
\ifthenelse{\boolean{showcomments}}
	{\newcommand{\nb}[3]{
		{\colorbox{#2}{\bfseries\sffamily\scriptsize\textcolor{white}{#1}}}
		{\textcolor{#2}{\sf\small$\blacktriangleright$\textit{#3}$\blacktriangleleft$}}}
	 }
	{\newcommand{\nb}[3]{}
	 }

\newcommand{\nadel}[1]{}  

\begin{document}

\title{Rethinking Training Data for Generating \\ Code Review Comments
}

\author{\IEEEauthorblockN{1\textsuperscript{st} Leonardo Centellas-Claros}
\IEEEauthorblockA{\textit{Department of Computer Science, School of Engineering} \\
\textit{Pontificia Universidad Católica de Chile}\\
Santiago, Chile \\
lcentellas@uc.cl}
\and
\IEEEauthorblockN{2\textsuperscript{nd} Estefania Pakarati-Cofre}
\IEEEauthorblockA{\textit{Department of Computer Science, School of Engineering} \\
\textit{Pontificia Universidad Católica de Chile}\\
Santiago, Chile \\
estefania.pakarati@uc.cl}
\and
\IEEEauthorblockN{3\textsuperscript{rd} Juan Pablo Sandoval Alcocer}
\IEEEauthorblockA{\textit{Department of Computer Science, School of Engineering} \\
\textit{Pontificia Universidad Católica de Chile}\\
Santiago, Chile \\
juanpablo.sandoval@uc.cl}
\and
\IEEEauthorblockN{4\textsuperscript{th} Diego Elias Costa}
\IEEEauthorblockA{\textit{REALISE Lab} \\
\textit{Concordia University}\\
Montreal, Canada \\
diego.costa@concordia.ca}
}

\maketitle

\begin{abstract}
Generating code review comments has become a prominent research direction in automated code review, commonly formulated as a text generation task over diff-comment pairs.
Despite advances in learning-based approaches, generated review comments are often generic, weakly grounded, or non-actionable.
Recent studies have also shown that review comment datasets contain noisy or unsuitable training instances, motivating LLM-based dataset cleaning approaches.
In this paper, we argue that problematic training instances are not homogeneous and that some limitations stem from deeper issues in the task formulation itself.
Through an empirical inspection of a widely used review comment dataset, we identify \emph{misaligned training pairs}: instances where the relationship between the code change and the review comment does not provide a reliable learning signal for generating actionable review feedback from localized inputs.
We derive a taxonomy of misalignment capturing three recurring sources: \emph{semantic ambiguity}, \emph{lack of actionability}, and \emph{context dependence}.
We further explore whether incorporating this taxonomy into LLM-based filtering improves the identification of problematic training instances, observing that detecting misaligned training pairs remains challenging.
Based on these observations, we argue that improving review comment generation requires more than dataset cleaning alone, motivating explicit validity criteria, richer contextual inputs, and evaluation practices aligned with review intent and actionability.
\end{abstract}


\section{Introduction}
Automated code review aims to support developers by either providing feedback similar to that produced during human code review \cite{TowardsACR} or transforming code into a reviewed version \cite{AutoTransform}.
With the increasing adoption of learning-based approaches and large language models, generating review comments has become a prominent research direction in software engineering.
Most existing approaches formulate this task as a supervised text generation problem over diff-comment pairs mined from pull request reviews, where localized representations of code changes, such as diff hunks or modified methods, are used to predict human-written review comments \cite{CodeReviewer,CommentFinder,AUGER}.

Despite substantial progress in modeling techniques, the effectiveness of automatically generated review comments remains limited \cite{StateOfTheArt}.
Prior studies report that generated comments are often \emph{generic, weakly grounded, or non-actionable} \cite{Experience}.
To address these limitations, most prior work has focused on improving model architectures, scaling learning approaches, or refining prompting strategies.

More recently, researchers have started questioning the quality of review comment datasets themselves.
Tufano et al.~\cite{StateOfTheArt} and Liu et al.~\cite{TooNoisy} reported that a substantial portion of review comment datasets correspond to noisy or unsuitable training instances.
Liu et al.~\cite{TooNoisy} further explored LLM-based dataset cleaning, showing that automatically identifying valid training instances remains challenging and that cleaning alone only partially improves review comment generation performance.
Taken together, these observations suggest that the limitations of review comment generation may stem not only from modeling challenges, but also from how the task itself is instantiated through current training datasets.

However, we argue that problematic training instances are not homogeneous.
While some correspond to conventional forms of dataset noise, others may reflect a deeper structural issue related to the task formulation itself.
We hypothesize that some review comments are legitimate review interactions whose interpretation depends on contextual information not available in localized input representations, leading to a \emph{structural misalignment} between review comments, training inputs, and the intended learning objective.

In this paper, we empirically investigate these problematic training instances through an inspection of the \cite{CodeReviewer} dataset.
We derive a taxonomy of \emph{misaligned training pairs}: instances where the relationship between the code change and the review comment does not provide a reliable learning signal for generating actionable review feedback from localized inputs.
Our taxonomy identifies three recurring sources of misalignment: \emph{semantic ambiguity}, \emph{lack of actionability}, and \emph{context dependence}.
We further investigate whether incorporating our taxonomy into LLM-based filtering can improve the identification of misaligned training pairs, observing that distinguishing these cases remains challenging even when explicit misalignment categories are provided.

Taken together, these observations suggest that improving review comment generation requires more than dataset cleaning: some problematic cases stem from deeper limitations in how the task is formulated through localized-input representations. Progress therefore requires \emph{explicit validity criteria} for training instances, \emph{richer contextual inputs} that expose review dependencies, and evaluation practices aligned with \emph{review intent and actionability} rather than surface-level textual similarity.

\section{Automated Review Comment Generation}

Code review serves as both a quality assurance mechanism and a means of knowledge sharing \cite{ModerCodeReview}, with reviewers providing feedback through natural-language comments that suggest improvements or identify issues. Automating this feedback has attracted increasing research attention \cite{literatureReview}.

Most existing approaches to review comment generation formulate the task as a supervised text generation problem. Training data is typically constructed by mining pull request reviews, where each instance consists of a code change paired with a corresponding review comment \cite{CodeReviewer}. Depending on the approach, the input is operationalized using different localized views of the change, including the entire code diff associated with the comment \cite{Experience,FineTuneLLM}, a diff hunk \cite{CodeReviewer}, or only the modified method after the change \cite{CommentFinder,TowardsACR}.

Given a localized representation of a code change as input, models are trained to generate review comments that resemble those written by human reviewers.
This formulation has been widely adopted across learning-based approaches, including encoder-decoder architectures \cite{AUGER} and, more recently, large language models \cite{FineTuneLLM}.

While intuitive, this formulation implicitly defines what it means to ``generate a review comment''.
By treating localized code representations as self-contained training instances, datasets assume that the available input is sufficient to infer review intent and actionability, and that all reviewer comments constitute appropriate generation targets.
As a result, the task is effectively reduced to reproducing the surface form of review comments, rather than modeling when and how actionable feedback should be provided.
This task definition has direct implications.
When relevant information lies outside the localized input, models are forced to rely on generic patterns observed in the data, which can yield fluent but weakly grounded comments and limit actionability.

\section{Empirical Observations on Review Comment Datasets}

Rather than evaluating model performance or dataset quality in isolation, we focus on characterizing misaligned training pairs in datasets for review comment generation and on understanding how these misalignments shape the learning objective induced by current task formulations.

\subsection{Methodology}

We follow a four-step methodology:

\begin{enumerate}
    \item \emph{Dataset selection.} To enable comparison with related studies such as \cite{TooNoisy}, we ground our empirical observations in a widely used public dataset for review comment generation, whose training split contains 117,739 diff-comment pairs \cite{CodeReviewer}. We selected this dataset because it is representative of current practice in automated code review \cite{StateOfTheArt} and has served as the basis for numerous studies, comparisons, and models since its publication~\cite{Experience,llamareviewer,ChatGPTACR}.
    
    \item \emph{Sample collection.}  As analyzing all training pairs is not feasible, we determined the sample size using standard random sampling with a 95\% confidence level and a 5\% margin of error, resulting in a required sample of 383 training pairs. These pairs were drawn uniformly at random and independently inspected by two reviewers. Each sampled pair consists of a code diff and its associated review comment.

    \item \emph{Instance screening.} 
    Two reviewers independently assessed whether each sampled diff-comment pair constituted actionable review feedback that a model would be expected to generate, having only received the diff as input. Both reviewers are postgraduate researchers in software engineering, with prior experience in code review and software quality analysis. Inter-rater reliability was substantial ($\kappa = 0.758$). Disagreements were subsequently discussed to reach a consensus, resulting in a final set of 184 misaligned training pairs.

    \item \emph{Taxonomy construction.} Following the screening step, we obtained a set of 184 training pairs identified as \emph{misaligned} with the objective of generating actionable review comments from change-local inputs. We based the taxonomy construction on an open card sorting approach, following the methodology proposed by \cite{LessIsMore}. For each training pair, reviewers assigned descriptive category labels capturing the underlying source of the misalignment, and recorded short notes to clarify category boundaries. As the analysis progressed, categories were iteratively revised, merged, or expanded.

\end{enumerate}

\begin{table*}[ht!]
\caption{Taxonomy of misaligned training pairs in review comment datasets.}
\label{tab:taxonomy}
\small
\centering
\begin{tabular}{p{3.2cm} p{0.5cm} p{13cm}}
\hline
\textbf{Category} & \textbf{Freq.} & \textbf{Description} \\
\hline

\multicolumn{3}{l}{\textbf{Semantic Ambiguity}} \\
\quad Unsure &
5 &
Expresses uncertainty about whether an issue exists, without proposing a concrete action. \\
\quad Unclear &
20 &
Intent is ambiguous or incomplete, preventing inference of a concrete action by the developer. \\

\hline
\multicolumn{3}{l}{\textbf{Lack of Actionability}} \\
\quad Inquiry &
56 &
Asks for clarification, but does not suggest or justify any modification. \\
\quad Comment &
9 &
Refers to the pull request without requesting changes or providing a clear purpose. \\
\quad \quad Chit-chat &
1 &
Casual or conversational remark with no technical relevance. \\
\quad \quad Remark &
32 &
Descriptive or acknowledging statement that does not imply action and often reflects self-commentary. \\

\hline
\multicolumn{3}{l}{\textbf{Context Dependence}} \\
\quad Context-dependent &
44 &
Although the comment may be useful, it is not possible for a model or reviewer to infer this comment from the information present in the diff alone. \\
\quad \quad Another change &
1 &
Refers to a change located in another diff of the same commit. \\
\quad \quad Other line &
14 &
Refers to code outside the displayed diff within the same file. \\
\quad Previous comment &
1 &
Explicitly refers to an earlier review comment. \\
\quad Broken URL &
1 &
Discusses a URL whose validity cannot be determined from the available data. \\

\hline
\end{tabular}
\end{table*}

\subsection{Results: Taxonomy of Misaligned Training Pairs}

Table \ref{tab:taxonomy} summarizes the resulting taxonomy of \emph{misaligned training pairs}, which groups these cases into three high-level categories: \emph{semantic ambiguity}, \emph{lack of actionability}, and \emph{context dependence}.
Rather than representing isolated noise, these categories capture systematic patterns that reflect different review-related intents or dependencies on information that is not available in the provided change-local input.

\begin{itemize}
    \item \emph{Semantic ambiguity (25/184).} This category includes comments whose intent is unclear or under-specified, preventing the inference of a concrete action. This category covers cases where reviewers explicitly express uncertainty about whether an issue exists, as well as comments whose phrasing is incomplete or vague. Although such comments are part of real review conversations, they do not provide a clear learning signal for generating actionable feedback grounded in the code.
    \item \emph{Lack of actionability (98/184).}  This category encompasses comments that serve conversational, coordinative, or descriptive purposes rather than proposing or justifying code changes. This includes inquiries that ask for clarification without suggesting a modification, as well as remarks and acknowledgments that comment on the pull request without implying any required action. While these comments are legitimate artifacts of the review process, they are far from the actionable review feedback expected of a model's output.
    \item \emph{Context dependence (61/184).}  This category captures comments that are potentially actionable, but whose correctness depends on information not present in the localized input. These include references to other changes within the same commit, code outside the displayed diff, prior review comments, or external artifacts such as URLs. For such instances, generating appropriate feedback requires access to additional context beyond the diff-comment representation. We argue that a model trained on these types of instances would learn to guess a review rather than to infer it from its input.
\end{itemize}

While some degree of noise is expected in large-scale datasets, these cases go beyond incidental dataset noise.
\textbf{Our analysis shows that 184 out of 383 sampled training pairs (48\%) are misaligned training pairs}, with lack of actionability and missing contextual information being the two most frequent sources of misalignment. This indicates that such cases constitute a substantial portion of the dataset rather than isolated anomalies.
This proportion is also in line with prior work reporting large numbers of problematic review-comment pairs, including 38\%~\cite{StateOfTheArt} and 25\% noisy instances~\cite{TooNoisy}. Note that, although related, our analysis goes beyond dataset noise and focuses on \textit{misalignment}.
A replication package with all experimental data is available online \cite{replication}.

\section{Revisiting LLMs as a Data Filter}

Liu et al.~\cite{TooNoisy} previously explored the use of LLMs to filter problematic review-comment pairs, reporting limited effectiveness. In this section, we revisit this idea to evaluate whether our taxonomy helps LLMs better distinguish aligned from misaligned training pairs, and whether the choice of model and input representation influences filtering performance.

\begin{figure}[t]
\centering
\scriptsize


\begin{tabular}{|p{0.92\columnwidth}|}
\hline
\multicolumn{1}{|c|}{\textbf{System Prompt}} \\[1mm]

\textbf{Role and Task} \\

Your task, as an experienced code reviewer is ...\\
<TASK\_INSTRUCTION> \\[1mm]

\rowcolor{yellow!20}You are an expert code reviewer evaluating whether a human-written review comment is suitable training data ...\\

\textbf{Definition} \\

<SUITABLE\_INSTANCE\_DEFINITION>
\\[1mm]

\textbf{Evaluation Criteria} \\

Your evaluation should be guided by the following criteria:
\begin{enumerate}
    \item Relevance to Code Change: ...
    \item Clarity and Constructiveness: ...
    \item Focus on Improvement: ...
\end{enumerate} \\
\\
\rowcolor{yellow!20} \textbf{Categories} \\
\rowcolor{yellow!20} Consider the following categories:\\
\rowcolor{yellow!20} \begin{enumerate}
    \item <CATEGORY\_DEFINITION\_1>
    \item <CATEGORY\_DEFINITION\_2>
    \item ...
\end{enumerate}
\\[1mm]
\end{tabular}

\begin{tabular}{|p{0.92\columnwidth}|}
\hline
\multicolumn{1}{|c|}{\textbf{User Prompt}} \\

Below is a code diff and review comment.\\
Please evaluate whether this comment is Valid or Noisy.\\[1mm]
\textbf{Context Code Changes:} \\
<CODE\_DIFF> \\[1mm]

\textbf{Input Review Comment:} \\
<REVIEW\_COMMENT> \\[1mm]
\textbf{Answer Format:} \\
valid or noisy\\[1mm] \hline
\end{tabular}

\caption{Prompt structure used in our study. Yellow-highlighted sections indicate the modifications introduced over the prompt of Liu et al.~\cite{TooNoisy}.}
\label{fig:prompt_template}

\end{figure}

\subsection{Methodology}
We follow a five-step methodology to evaluate whether taxonomy-guided prompts can improve the filtering of unsuitable review-comment pairs for automated code review generation.

\begin{enumerate}

    \item \emph{Datasets.} We conducted experiments using two manually annotated datasets. First, we used the dataset released by Liu et al.~\cite{TooNoisy}, which contains 270 review-comment pairs sampled from the CodeReviewer dataset and annotated according to the suitability of review comments for automated code review generation. Second, we used our manually curated dataset introduced in this paper, which contains 383 review-comment pairs labeled as either \textit{misaligned} or \textit{aligned}.
   
    \item \emph{LLMs under Study.} We evaluate \emph{GPT-3.5-turbo}, matching the original setup on \cite{TooNoisy}, and \emph{GPT-4o-mini}, a more recent lightweight model, to test whether newer LLMs better identify unsuitable pairs, particularly with taxonomy-derived prompts.
    
    \item \emph{System Prompt (Task Instruction).} 
    We evaluated three prompting strategies that progressively extend the information provided through the \textit{system prompt} to guide the LLM during the filtering task.
    
    \begin{itemize}
    
        \item $SP_{D}$ (\textit{Definition}): prompt based on Liu et al. \cite{TooNoisy} that includes only the reviewer role description together with the definitions of valid and noisy review comments used during manual annotation.
    
        \item $SP_{DC}$ (\textit{Definition+Criteria}): extension of $SP_{D}$, also developed by Liu et al. \cite{TooNoisy}, that additionally incorporates evaluation criteria and explicit guidance rules to help the model assess the relevance and quality of review comments.
    
        \item $SP_{DCT}$ (\textit{Definition+Criteria+Taxonomy}): our proposed prompt, based on $SP_{DC}$ by incorporating taxonomy-derived categories and criteria related to contextual alignment, actionability, and observable relation with the provided code diff.
    
    \end{itemize}
    Figure \ref{fig:prompt_template} shows the prompt structure proposed originally by Liu et al. and highlights in yellow the changes or additions we made to it. In all configurations, the \textit{task instruction} asks the model to determine whether the provided sample is Valid or Noisy. Our prompt adds to this framing by asking whether the pair is a suitable training instance for automated code review generation.

    \item \emph{User Prompt Representations.} 
    We evaluated two \textit{user prompt} input representations that differ in the contextual information provided to the LLM.
    
    \begin{itemize}
    
        \item $UP_{NL}$: includes only the natural language review comment.
    
        \item $UP_{NL+Diff}$: includes both the review comment and the associated code diff.
    
    \end{itemize}
    \item \emph{Procedure and Metrics} 
    We evaluated all possible combinations of system prompting strategies (3), LLMs (2), and user prompt representations (2), resulting in a total of 12 experimental configurations. For each configuration, the corresponding prompt setup was executed over the full dataset, and the LLM classified each review-comment pair as either a suitable or unsuitable training instance for automated code review generation. We additionally included a random baseline to test whether the prompt-based configurations outperformed chance-level labeling.
    
    We then compared the predicted labels against the ground truth annotations and computed precision, recall, and F1-score for both classes to evaluate how effectively each configuration distinguishes suitable from unsuitable review-comment pairs.

\end{enumerate}

\begin{table*}[ht!]
\centering
\caption{Results across datasets and configurations.}
\label{tab:results}
\resizebox{\textwidth}{!}{
\begin{tabular}{lll|cccc|cccc|cccc|cccc}
\toprule
\textbf{Prompt} & \textbf{GPT} & \textbf{Input} & \multicolumn{15}{c}{\textbf{Datasets}} & \\
\hline
 & & & \multicolumn{8}{c|}{\textbf{Liu et al. Dataset \cite{TooNoisy}}} & \multicolumn{8}{c}{\textbf{Our Dataset}} \\
 & & & \multicolumn{4}{c|}{Valid (172)} & \multicolumn{4}{c|}{Noisy (98)} & \multicolumn{4}{c|}{Valid (199)} & \multicolumn{4}{c}{Noisy (184)} \\

  &  &  & Prec & Rec & F1 & \# & Prec & Rec & F1 & \# & Prec & Rec & F1 & \# & Prec & Rec & F1 & \# \\ \hline
  \multicolumn{3}{l|}{Random} & 0.62 & 0.46 & 0.53 & 128 & 0.35 & 0.50 & 0.41 & 142 & 0.55 & 0.55 & 0.55 & 200 & 0.51 & 0.51 & 0.51 & 183 \\
\hline
  $SP_{D}$ & 3.5 & $UP_{NL}$ & 0.85 & 0.44 & \cellcolor{green!25}0.58 & 89 & 0.47 & 0.87 & \cellcolor{green!25}0.61 & 181 & 0.76 & 0.39 & 0.52 & 102 & 0.57 & 0.87 & \cellcolor{green!25}0.69 & 281 \\
  $SP_{D}$ & 3.5 & $UP_{NL+Diff}$ & 0.74 & 0.61 & \cellcolor{green!25}0.67 & 141 & 0.48 & 0.63 & \cellcolor{green!25}0.55 & 129 & 0.58 & 0.59 & \cellcolor{green!25}0.59 & 200 & 0.55 & 0.55 & \cellcolor{green!25}0.55 & 183 \\
  $SP_{D}$ & 4o-mini & $UP_{NL}$ & 0.94 & 0.17 & 0.29 & 32 & 0.40 & 0.98 & \cellcolor{green!25}0.57 & 238 & 0.87 & 0.20 & 0.32 & 45 & 0.53 & 0.97 & \cellcolor{green!25}0.68 & 338 \\
  $SP_{D}$ & 4o-mini & $UP_{NL+Diff}$ & 0.94 & 0.09 & 0.17 & 17 & 0.38 & 0.99 & \cellcolor{green!25}0.55 & 253 & 0.76 & 0.13 & 0.22 & 34 & 0.50 & 0.96 & \cellcolor{green!25}0.66 & 349 \\
\hline
  $SP_{DC}$ & 3.5 & $UP_{NL}$ & 0.82 & 0.63 & \cellcolor{green!25}0.71 & 133 & 0.54 & 0.76 & \cellcolor{green!25}0.63 & 137 & 0.71 & 0.60 & \cellcolor{green!25}0.65 & 167 & 0.63 & 0.74 & \cellcolor{green!25}0.68 & 216 \\
  $SP_{DC}$ & 3.5 & $UP_{NL+Diff}$ & 0.72 & 0.57 & \cellcolor{green!25}0.64 & 136 & 0.45 & 0.61 & \cellcolor{green!25}0.52 & 134 & 0.60 & 0.55 & \cellcolor{green!25}0.58 & 183 & 0.56 & 0.60 & \cellcolor{green!25}0.58 & 200 \\
  $SP_{DC}$ & 4o-mini & $UP_{NL}$ & 0.92 & 0.26 & 0.41 & 49 & 0.43 & 0.96 & \cellcolor{green!25}0.59 & 221 & 0.85 & 0.27 & 0.41 & 62 & 0.55 & 0.95 & \cellcolor{green!25}0.69 & 321 \\
  $SP_{DC}$ & 4o-mini & $UP_{NL+Diff}$ & 1.00 & 0.09 & 0.17 & 16 & 0.39 & 1.00 & \cellcolor{green!25}0.56 & 254 & 0.79 & 0.12 & 0.20 & 29 & 0.50 & 0.97 & \cellcolor{green!25}0.66 & 354 \\
\hline
  $SP_{DCT}$ & 3.5 & $UP_{NL}$ & 0.79 & 0.61 & \cellcolor{green!25}0.69 & 133 & 0.51 & 0.71 & \cellcolor{green!25}0.60 & 137 & 0.65 & 0.61 & \cellcolor{green!25}0.63 & 186 & 0.60 & 0.65 & \cellcolor{green!25}0.62 & 197 \\
  $SP_{DCT}$ & 3.5 & $UP_{NL+Diff}$ & 0.74 & 0.66 & \cellcolor{green!25}0.70 & 155 & 0.50 & 0.58 & \cellcolor{green!25}0.54 & 115 & 0.56 & 0.59 & \cellcolor{green!25}0.57 & 210 & 0.53 & 0.49 & 0.51 & 173 \\
  $SP_{DCT}$ & 4o-mini & $UP_{NL}$ & 0.87 & 0.40 & \cellcolor{green!25}0.55 & 79 & 0.46 & 0.90 & \cellcolor{green!25}0.61 & 191 & 0.81 & 0.41 & 0.55 & 101 & 0.59 & 0.90 & \cellcolor{green!25}0.71 & 282 \\
  $SP_{DCT}$ & 4o-mini & $UP_{NL+Diff}$ & 0.87 & 0.15 & 0.26 & 30 & 0.39 & 0.96 & \cellcolor{green!25}0.56 & 240 & 0.77 & 0.14 & 0.23 & 35 & 0.51 & 0.96 & \cellcolor{green!25}0.66 & 348 \\
\bottomrule \\
\end{tabular}
}
\begin{minipage}{\textwidth}
\footnotesize
\textit{Note:} $SP_{D}$ and $SP_{DC}$ denote the prompts from Liu et al.~\cite{TooNoisy}, while $SP_{DCT}$ extends the original prompt with taxonomy-derived category definitions. $UP_{NL}$ uses only the review comment, whereas $UP_{NL+Diff}$ also includes the code diff. Green cells indicate F1 scores above the random baseline.
\end{minipage}
\end{table*}

\subsection{Results}

Table~\ref{tab:results} reports the results across the evaluated prompting configurations. $UP_{NL}$ and $UP_{NL+Diff}$ produced similar results, with $UP_{NL}$ slightly more consistent across datasets. GPT-3.5-turbo consistently outperformed GPT-4o-mini. In several configurations, GPT-4o-mini classified most samples as unsuitable instances, resulting in low recall for the Valid class.

Differences between our taxonomy-guided prompt ($SP_{DCT}$) and the prompts derived from Liu et al.~\cite{TooNoisy} ($SP_{D}$ and $SP_{DC}$) were relatively small. Although $SP_{DCT}$ achieved slightly better results in some configurations, the improvements were marginal and inconsistent across datasets.

Overall, \textbf{adding taxonomy-derived categories to the prompt did not substantially improve filtering effectiveness}. Still, LLM-based filtering captured some signal beyond chance, with GPT-3.5-turbo achieving F1 scores between 0.58 and 0.71 for the Valid class across the best-performing configurations. These results suggest that more effective filtering may require clearer validity criteria, richer contextual information, and approaches beyond zero-shot prompting.

\section{Discussion: From Noise to Misalignment}

Our empirical observations reveal limitations beyond isolated data quality issues. A substantial portion of training instances encode intents misaligned with the goal of producing actionable, change-grounded feedback. Rather than reflecting low-quality data, these instances expose a structural mismatch between the intended task and its instantiation in commonly used datasets and change-local inputs.

\myparagraph{Beyond noisy data}
At first glance, ambiguous or non-actionable comments may appear to be simple noise that could be addressed through dataset cleaning \cite{StateOfTheArt}.
However, our categories capture systematic patterns that correspond to legitimate artifacts of real-world code review, including clarification, coordination, and conversational interaction.
Treating these instances as noise obscures a deeper issue: the learning objective induced by current datasets conflates actionable review feedback with broader review discourse. Our findings suggest that current review comment datasets capture interactions serving different purposes in human code review, although only a subset naturally aligns with generating actionable review comments from localized code changes.


\myparagraph{Task misalignment through training data}
By mixing actionable feedback with comments that serve other purposes, datasets implicitly redefine what models are expected to learn.
As a result, models are rewarded for producing fluent, review-like text even when a training instance does not support actionable feedback grounded in the provided input.
This helps explain why generated comments can be plausible yet unhelpful: improvements in fluency or stylistic similarity do not necessarily translate into actionable review behavior.

\myparagraph{Exploring automatic data filtering}
As an exploratory analysis, we investigated whether noisy training instances could be automatically filtered using a large language model. The results suggest that identifying valid training instances is not trivial and requires explicit validity criteria and sensitivity to review intent and missing context. Future work could explore more sophisticated prompting techniques or richer LLM-based configurations, as the straightforward approach we tested leaves considerable room for improvement.

\myparagraph{Structural limits of change-local inputs}
Context-dependent instances represent a substantial fraction of the misaligned training pairs in our taxonomy. In these cases, the review comment cannot be inferred from change-local inputs alone, as essential information lies outside the diff, such as references to other changes, code outside the snippet, or prior review comments. For these instances, misclassification reflects missing context rather than a failure of the model. Addressing this limitation would require input representations that extend beyond strictly change-local inputs. These context-dependent instances may also provide a useful benchmark for evaluating approaches that retrieve additional context before generating feedback, including emerging agentic review systems.

\section{Rethinking Data and Inputs for Review Comment Generation}

The observations derived from our dataset analysis suggest that advancing review comment generation requires rethinking how the task is defined through data and input representations.
Rather than assuming that improvements in model capacity will compensate for misaligned or incomplete data, we argue for a data- and input-centric perspective in which the learning objective is made explicit and better aligned with real review practices.

\myparagraph{Validity as a first-class property}
Current datasets implicitly treat all diff-comment pairs as equally valid training instances, an assumption our taxonomy shows does not hold in practice. Future datasets should explicitly define what constitutes a valid instance, distinguishing actionable review feedback from other interactions such as discussion, coordination, or clarification. Making instance validity explicit would help ensure models are trained on signals aligned with the intended task, rather than reproducing heterogeneous review behavior.

\myparagraph{Context-aware input representations}
The prevalence of context-dependent comments highlights the limitations of diff-only inputs.
To support actionable review behavior, input representations should incorporate richer contextual signals, such as review-thread history, references to non-local code, project-specific conventions, or signals from continuous integration pipelines.
Importantly, context should not be treated as an optional enhancement, but as a necessary component for a substantial class of review interactions.

\myparagraph{Explicit handling of missing context}
Even with richer inputs, there will remain situations in which the available information is insufficient to generate meaningful review feedback.
Rather than forcing models to produce a comment in all cases, future approaches should explicitly model the possibility of insufficient context.
Enabling models to abstain, defer, or request additional information may be preferable to generating fluent but weakly grounded comments, and better reflects realistic review behavior.

\section*{Acknowledgment}
Leonardo Centellas-Claros is supported by the Chilean National Agency for Research and Development (ANID)/Scholarship Program/DOCTORADO NACIONAL/ 2024-21240734 and National Center for Artificial Intelligence (CENIA), Grant/Award Number: BASAL, ANID, FB210017.
\

\bibliographystyle{IEEEtran}
\bibliography{references}

\end{document}